\documentclass[fleqn,11pt]{article}

\usepackage{amsmath,latexsym,amssymb,cite}
\usepackage{epsf,graphicx}

\usepackage{amsfonts,amssymb,cite}
\usepackage{epsf,graphicx}

\newcommand{\bls}[1]{\renewcommand{\baselinestretch}{#1}}

\def\noi{\noindent}

\newcommand{\Title}[1]{\noi {{\Large\bf #1}}\\[1ex]}

\def\Aunames#1{\noi{\bf #1}}
\def\auth#1{${}^{#1}$}
\def\Addresses#1{\medskip\noi \protect
	\begin{description}\itemsep -3pt {\it #1} \end{description}}
\def\addr#1#2{\item[${}^{#1}$]{\it #2}}

\newcommand{\Rec}[1]{\noi {\it Received #1} \\}

\newcommand{\Abstract}[1]{\vskip 2mm \begin{center}
        \parbox{16.4cm}{\small\noi #1} \end{center}\medskip}

\newcommand{\foox}[2]{\footnotetext[#1]{#2}\addtocounter{footnote}{1}}
\def\email#1#2{\footnotetext[#1]{e-mail: #2}\addtocounter{footnote}{1}}

\def\Plenary{\foox 1 {Based on a plenary talk given at the 11th International 
	   Conference on Gravitation, Astrophysics and Cosmology 
	   of Asia-Pacific Countries (ICGAC-11),
	   October 1--5, 2013, Almaty, Kazakhstan}}

\def\nq{\hspace*{-1em}}
\def\nqq{\hspace*{-2em}}
\def\nhq{\hspace*{-0.5em}}

\def\cm{\hspace*{1cm}}

\def\Jl#1#2{#1 {\bf #2},\ }

\def\ApJ#1 {\Jl{Astroph. J.}{#1}}
\def\CQG#1 {\Jl{Class. Quantum Grav.}{#1}}
\def\DAN#1 {\Jl{Dokl. AN SSSR}{#1}}
\def\GC#1 {\Jl{Grav. Cosmol.}{#1}}
\def\GRG#1 {\Jl{Gen. Rel. Grav.}{#1}}
\def\JETF#1 {\Jl{Zh. Eksp. Teor. Fiz.}{#1}}
\def\JETP#1 {\Jl{Sov. Phys. JETP}{#1}}
\def\JHEP#1 {\Jl{JHEP}{#1}}
\def\JMP#1 {\Jl{J. Math. Phys.}{#1}}
\def\NPB#1 {\Jl{Nucl. Phys. B}{#1}}
\def\NP#1 {\Jl{Nucl. Phys.}{#1}}
\def\PLA#1 {\Jl{Phys. Lett. A}{#1}}
\def\PLB#1 {\Jl{Phys. Lett. B}{#1}}
\def\PRD#1 {\Jl{Phys. Rev. D}{#1}}
\def\PRL#1 {\Jl{Phys. Rev. Lett.}{#1}}

\def\al{&\nhq}
\def\lal{&&\nqq {}}

\def\beq{\begin{equation}}
\def\eeq{\end{equation}}
\def\bear{\begin{eqnarray}}
\def\bearr{\begin{eqnarray} \lal}
\def\ear{\end{eqnarray}}
\def\earn{\nonumber \end{eqnarray}}

\def\nnv{\nonumber\\[5pt] {}}
\def\nnn{\nonumber\\ \lal }

\def\yyy{\\[5pt] \lal }
\def\eql{\al =\al}

\def\tst{\textstyle}

\def\fract#1#2{{\tst\frac{#1}{#2}}}

\def\half{{\fract{1}{2}}}

\def\e{{\,\rm e}}
\def\d{\partial}

\def\sign{\mathop{\rm sign}\nolimits}
\def\diag{\mathop{\rm diag}\nolimits}

\def\const{{\rm const}}
\def\eps{\varepsilon}

\newcommand{\vars}[1]{\left\{\begin{array}{ll}#1\end{array}\right.}

\def\R{{\mathbb R}}

\def\Rs{{}_s R{}}
\def\Ro{{}_\omega R{}}
\def\Gs{{}_s G{}}
\def\Go{{}_\omega G{}}

\def\mn{_{\mu\nu}}
\def\MN{^{\mu\nu}}
\def\mN{_\mu^\nu}

\def\GR{general relativity}

\def\ssph{static, spherically symmetric}
\def\cy{cylindrical}
\def\cyl{cylindrically symmetric}
\def\scyl{static, cylindrically symmetric}
\def\sax{static, axially symmetric}

\def\bh{black hole}

\def\wh{wormhole}
\def\whs{wormholes}
\def\asflat{asymptotically flat}

\bls{0.99}
\begin{document}
\twocolumn[

\Title{Cylindrically and axially symmetric wormholes. Throats in vacuum? }

\Aunames{K. A. Bronnikov\auth{a,b,c,2} and M. V. Skvortsova\auth{a,3}}

\Addresses{
\addr a {Center for Gravitation and Fundam. Metrology, VNIIMS,
   	Ozyornaya ul. 46, Moscow 119361, Russia;}
\addr b  {Institute of Gravitation and
   	Cosmology, PFUR, ul. Miklukho-Maklaya 6, Moscow 117198, Russia;}
\addr c {I. Kant Baltic Federal University, ul. Al. Nevskogo 14,
	Kaliningrad 236041, Russia}
	   }

\Rec{April 15, 2014}

\Abstract
  {This brief review discusses the existence conditions of \wh\ throats
  and \whs\ as global configurations in \GR\ under the assumptions of \cy\
  and axial symmetries. It is pointed out, in particular, that \wh\ throats
  can exist in \scyl\ space-times under slightly different conditions as
  compared with spherical symmetry. In \cyl\ space-time with rotation,
  throats can exist in the presence of ordinary matter or even in vacuum;
  however, there are substantial difficulties in obtaining \asflat\ \wh\
  configurations without exotic matter: such examples are yet to be found.
  Some features of interest are discussed in \sax\ configurations, including
  wormholes with singular rings and wrongly seeming regular \wh\ throats in
  the Zipoy-Voorhees vacuum space-time.
  }

]  
\Plenary
\email 2 {kb20@yandex.ru}
\email 3 {milenas577@mail.ru}

\section{Introduction}

  Wormholes, the hypothetic narrow ``bridges'' or ``tunnels'' connecting
  different large or infinite regions of space-time or even different
  Universes, have become a subject of active discussion in the recent
  decades. Their possible existence can lead to physical effects of great
  interest, such as realizable time machines or shortcuts between distant
  parts of the Universe, in particular, across \bh\ horizons
  \cite{thorne, viss-book, ws_book}. Unusual observable effects can be
  predicted if \whs\ exist on astrophysical scales of lengths and times
  \cite{sha, astro, kir-sa1, kir-sa2}.

  As is well known, the existence of a static \wh\ geometry in the framework
  of \GR\ requires the presence of ``exotic'', or phantom matter, that is,
  matter violating the null energy condition (NEC), at least in a
  neighborhood of the throat \cite{thorne,viss-book,HV97,ws_book}. This
  conclusion, however, rests on the assumption that the throat is a compact
  2D surface, having a finite (minimum) area \cite{HV97}. In other words, a
  \wh\ entrance looks from outside as a local object like a star or a \bh.

  Since macroscopic exotic matter has not been observed in laboratory or
  in the Universe (except for the possible phantom dark energy), it is
  natural to try to   obtain phantom-free\footnote
	{That is, respecting the NEC, and with nonnegative matter density}
  \whs\ (or at least throats) by abandoning some of
  the assumptions of the Hochberg-Visser (HV) theorem \cite{HV97}. One of
  the ways is to reject compactness and to consider, as the simplest
  assumption, \cy\ symmetry. Then, instead of starlike
  structures, we deal with objects infinitely extended along a
  certain direction, like cosmic strings. One can also consider nonstatic,
  rotating configurations, which can be done again in the
  framework of \cy\ symmetry.
  Besides, some well-known \sax\ space-times possess \wh\ properties
  \cite{zipoy, BFab97}, but all of them contain singular rings,
  violating the regularity requirement of the HV theorem. We here briefly
  describe all such space-times.

  In addition, we discuss a somewhat unexpected phenomenon in another
  family of vacuum \sax\ space-times, namely, a branch of the Zipoy-Voorhees
  class of solutions \cite{zipoy, voorhees}: a seeming existence of regular
  2-surfaces of minimum area (i.e., throats) contrary to the HV theorem. One
  could even suspect that there is a loophole in the conditions of the HV
  theorem, by analogy with the ``topological censorship'' theorem
  \cite{kras13}. It turns out, however, that the ``suspicious'' surfaces are
  not minimal, and the HV theorem works quite well in this case.

\section{Static \cy\ \whs}

  In \ssph\ space-times with the general metric
\bearr     \nhq                                                \label{ds-sph}
    ds^2 = A(u) dt^2 - A(u) du^2
\nnn \cm
	- r^2(u) (d\theta^2 + \sin^2 \theta d\varphi^2)
\ear
  ($u$ being an arbitrary admissible radial coordinate), we say
  that there is a \wh\ geometry if at some $u=u_0$ the function $r(u)$ has a
  regular minimum $r(u_0) > 0$ (which is then called a throat) and, on both
  sides of this minimum, $r(u)$ grows to much larger values than $r(u_0)$.
  It is supposed that, at least in some range of $u$ containing $u_0$, the
  function $A(u)$ is smooth, finite and positive, which guarantees
  regularity and absence of horizons. The stress-energy tensor (SET) of
  matter compatible with this symmetry has the form $T\mN = \diag(\rho,
  -p_r, -p_\bot, -p_\bot)$, where $\rho$ is the matter density and $p_r$ and
  $p_\bot$ are pressures in the radial and angular directions, respectively.
  Then, one of the Einstein equations has the form (in units with $c=G=1$)
\beq                                                     \label{01-sph}
  	2 A r''/r = -8\pi (\rho + p_r),
\eeq
  where the prime stands for $d/du$. A minimum of $r(u)$ requires $r''> 0$,
  hence $\rho + p_r < 0$ at the throat,\footnote
	{For simplicity, we consider generic minima at which, e.g., $r''>0$.
	 At more special minima one can have $r''=0$ at the throat but then
	 inevitably $r'' > 0$ in its neighborhood, and similarly in other
	 cases.}
  which means NEC violation. It is the simplest manifestation of the HV
  theorem.

  In the \scyl\ metric
\bearr                                                     \label{ds-cy}
    ds^2 = \e^{2\gamma(u)} dt^2 - \e^{2\alpha(u)} du^2
\nnn\cm
	- \e^{2\mu (u)} dz^2 - \e^{2\beta(u)} d\varphi^2
\ear
  ($u$ is any admissible radial coordinate, $z\in \R$ and
  $\varphi \in [0,\ 2\pi]$ are the longitudinal and angular ones), there
  are two reasonable analogues of the squared spherical radius $r^2$ in
  (\ref{ds-sph}): the squared circular radius $r^2(u) = \e^{\beta(u)}$ and
  the area function $a(u) = \e^{\mu+\beta}$. Accordingly \cite{BLem09},
  \cy\ \wh\ throats can be defined as regular minima of $r(u)$ (to be called
  $r$-throats) or $a(u)$ (to be called $a$-throats).

  Then \cite{BLem09}, taking the most general SET compatible with
  (\ref{ds-cy}), $T\mN = \diag (\rho, -p_r, -p_z, -p_\varphi)$,
  and using proper combinations of the Einstein equations, we arrive at
  the following conditions that must hold at and near the throat:
\bearr                                                    \label{r-th}
    	\rho - p_r - p_z + p_\varphi < 0  \quad (r{\rm -throat}),
\yyy                                                      \label{a-th}
    	\rho < p_r \leq 0\ \ \ \cm (a{\rm -throat}).
\ear
  Thus an $r$-throat does not necessarily require violation of any of the
  standard energy conditions, whereas at and near an $a$-throat there is
  always a region with negative energy density $\rho$.

  All that concerned only local conditions at the throat. However, to obtain
  a \wh\ observable as a stringlike source of gravity from an otherwise very
  weakly curved or even flat environment, we should require the existence of
  two spatial infinities on both sides of the throat, i.e., such values
  $u = u_{\pm\infty}$ that $r = \e^\beta \to \infty$ and the metric is either
  flat or corresponds to the gravitational field of a cosmic string. In other
  words, at both ends of the $u$ range we must have $\beta\to
  \infty$ and finite limits of $\gamma$ and $\mu$. Hence such a
  configuration should contain {\it both\/} $r$- and $a$-throats (not
  necessarily coinciding), and the latter inevitably requires $\rho < 0$ by
  (\ref{a-th}). We thus obtain a no-go theorem \cite{BLem09}:

  {\it A \scyl\ \wh\ with two flat or stringlike asymptotic regions cannot
  exist if the energy density $\rho$ is everywhere nonnegative.}

  Accordingly, all numerous examples of static phantom-free \cy\ \whs\
  (\cite{BLem09} and references therein) are not \asflat\ and contain only
  $r$-throats.

\section{Rotating \cy\ \\ \whs}

  A stationary \cyl\ metric with rotation can be written as
\bearr                                                    \label{ds-rot}
       ds^2 = \e^{2\gamma(u)}[dt - E(u)\e^{-2\gamma(u)}\, d\varphi]^2
\nnn \cm
       - \e^{2\alpha(u)}du^2 + \e^{2\mu(u)}dz^2 + \e^{2\beta(u)}d\varphi^2,
\ear
  where the second line gives the three-dimensional line element. The
  definitions of $r$- and $a$-throats are the same as before in terms of
  $\e^\beta$ and $\e^\mu$. A new feature as compared to the static case is
  the emergence of a vortex gravitational field described as a 4-curl of the
  tetrad $e^\mu_a$: its kinematic characteristic is the angular velocity
  of tetrad rotation \cite{kr2}
\beq                                                      \label{def-o}
      \omega^\mu = \half \eps^{\mu\nu\rho\sigma} e_{m\nu} e^m_{\rho;\sigma},
\eeq
  where the Latin letters $m, n, \ldots$ stand for Lorentz indices. It turns
  out that in the gauge $\alpha = \mu$ the vortex $\omega
  = \sqrt{\omega_\alpha \omega^\alpha}$ is \cite{BLem13, kr3, kr4}
\beq                                                  \label{o-def}
     \omega = \half (E\e^{-2\gamma})' \e^{\gamma-\beta-\mu}.
\eeq

  Furthermore, the off-diagonal component of the Ricci tensor $R_0^3$ is, in
  the same gauge, given by
\beq                                                   \label{R_30}
	\sqrt{-g} R^3_0 = -(\omega \e^{2\gamma+\mu})',
	\ \ \ g := \det (g\mn).
\eeq
  Assuming that our rotating reference frame is comoving to the matter
  source of gravity, that is, the azimuthal flow $T^3_0 =0$, we find from
  $R^3_0 =0$ that
\beq       	      					\label{omega}
	\omega = \omega_0 \e^{-\mu-2\gamma}, \cm \omega_0 = \const.
\eeq

  As a result \cite{BLem13}, the diagonal components of the Ricci tensor
  $R\mN$ can be written as the corresponding components $\Rs\mN$ for
  the static metric (\ref{ds-cy}) plus the $\omega$-dependent addition
\beq
         \Ro\mN = \omega^2 \diag (-2,\ 2,\ 0,\ 2),  \label{Ric-o}
\eeq
  The Einstein tensor $G\mN = R\mN - \half \delta\mN R$ splits
  in a similar manner, $G\mN = \Gs\mN + \Go\mN$, where
\beq
	\Go\mN = \omega^2 \diag (-3,\ 1,\ -1,\ 1).        \label{Ein-o}
\eeq
  One can check that the tensors $\Gs\mN$ and $\Go\mN$ (each separately)
  satisfy the ``conservation law'' $\nabla_\alpha G^\alpha_\mu =0$ with
  respect to the static metric (\ref{ds-cy}).

  Then, according to the Einstein equations $G\mN = - 8\pi T\mN$,
  the tensor $\Go\mN/(8\pi)$ behaves as an additional SET with very exotic
  properties (thus, the effective energy density is
  $-3\omega^2/(8\pi) < 0$), acting in the auxiliary metric (\ref{ds-cy}).
  In its presence, the existence conditions for \wh\ throats read
\bearr   \nq                                                 \label{r-thr}
    	\rho - p_r - p_z + p_\varphi < \omega^2/(4\pi)
			\ \ (r{\rm -throat}),
\yyy     \nq                                                 \label{a-thr}
    	\rho - p_r < \omega^2/(4\pi), \ \ p_r \leq \omega^2/(8\pi)
	 		\ (a{\rm -throat}).
\ear
  It is much easier to fulfil them than in the static case, as
  confirmed by a number of examples \cite{kr3, kr4, BLem13}.

  For example, if matter is represented by a massless, minimally coupled
  scalar field with the Lagrangian $L_s = g\MN \d_\mu\phi\d_\nu\phi$,
  the Einstein-scalar equations lead to the following solution with the
  metric (\ref{ds-rot}):
\bear
       \e^{2\beta} \eql \frac{\e^{2hu}}{2\omega_0 s(k,u)},\cm\ \qquad
       \e^{2\mu} = \e^{-2mu},
\nnv                                                      \label{sol1}
       \e^{2\gamma} \eql 2\omega_0 s(k,u) \e^{2hu},\cm
       \e^{2\alpha} = \e^{(4h-2m)u},
\nnv
       \omega \eql \frac{\e^{mu - 2hu}} {2s(k,u)},\cm  \phi = Cu
\nnv
       E \eql \e^{2hu}[E_0 s(k,u) - s'(k,u)],
\ear
  where the function $s(k,u)$ is defined as
\[
	s(k,u) = \vars{
		k^{-1} \sinh ku, & k > 0,\  u \in \R_+;\\
			      u, & k=0, \  u \in \R_+; \\
		 k^{-1} \sin ku, & k<0, \  0 < u < \pi/|k|;
		}
\]
  $\omega_0,\ E_0,\ h,\ k,\ m,\ C$ are integration constants obeying the
  relation
\beq                                                      \label{int1}
	k^2 \sign k = 4(h^2 - 2hm) - 32 \pi C^2.
\eeq
  This solution is written using the harmonic coordinate $u$ corresponding
  to the gauge condition $\alpha = \beta + \gamma + \mu$. If the scalar
  charge $C$ is zero, it is a vacuum solution.

  In all branches of the solution, $r\to \infty$ and $\e^\gamma \to 0$ as
  $u\to 0$. In the same limit, the vortex $\omega \to \infty$, indicating a
  singularity. At the other end of the $u$ range, the situation is more
  diverse:

\medskip\noi
{\bf 1. $k > 0$.} At large $u$, $\e^{2\beta} \sim \e^{(2h-k)u}$ and
  $\e^{2\gamma} \sim \e^{(2h+k)u}$, hence a \wh\ with an $r$-throat exists
  if $0 < k < 2h$; we also have $\e^\gamma \to \infty$ at large $u$.
  A \wh\ with an $a$-throat exists if $0 < k < 2(h-m)$.

\medskip\noi
{\bf 2. $k = 0$.} At large $u$ we have $\e^{2\beta} \sim u^{-1}\e^{2hu}$ and
  $\e^{2\gamma} \sim u \e^{2hu}$, hence we have a \wh\ geometry with an
  $r$-throat if $h >0$ and with an $a$-throat if $h-m > 0$.
  In addition, $\e^\gamma \to \infty$ at large $u$.

\medskip\noi
{\bf 3. $k < 0$.} A \wh\ geometry (with both kinds of throats) is described
  by all solutions with $k < 0$. At both ends, $\e^\beta \to \infty$
  and $\e^{\gamma}\to 0$, while $\e^{\beta + \gamma}$ and $\e^\mu$ remain
  finite, and $ \omega \sim \e^{-2\gamma} \to \infty$.

\medskip
  We conclude that rotating \cy\ vacuum and scalar-vacuum space-times are
  quite generically of \wh\ nature, and, in particular, the question asked
  in the title is answered ``yes''.

  Though, none of these rotating \wh\ solutions are \asflat. An attempt to
  remove this shortcoming was made in \cite{BLem13} by using a cut-and-paste
  procedure: on both sides of the throat, a \wh\ solution is matched to a
  properly chosen region of flat space (in a rotating reference frame) at
  some surfaces $\Sigma_-$ and $\Sigma_+$. It was shown, however, that if
  the throat region is described by the above vacuum or scalar-vacuum
  solutions, one or both thin shells appearing on $\Sigma_-$ and $\Sigma_+$
  inevitably violate the NEC. Thus, although phantom-free rotating \wh\
  solutions are easily found, exotic matter is still necessary for obtaining
  asymptotic flatness.

\section {Static, axially symmetric\\ \whs\ and throats}

  Consider the Zipoy-Voorhees vacuum \sax\ space-times \cite{zipoy,voorhees}
  with the metric
\bearr
    ds^2 = \e^{2\gamma} dt^2 - L^2 \e^{2\gamma + 2\eta}(x^2 + \eps y^2)
\nnn \cm \cm
	  \times \biggl(\frac{dx^2}{x^2+\eps} + \frac{dy^2}{1-y^2}\biggr)
\nnn \cm                                                    \label{ds-ax}
	  - L^2 \e^{-2\gamma} (x^2 + \eps)(1 - y^2)d\phi^2,
\ear
  where $L$ is a certain (arbitrary) length scale, $\eps = 0, \pm 1$
  designates three branches of the solution; $y \in (-1, +1)$ is a
  coordinate resembling latitude in spherical symmetry, $\phi\in [0, 2\pi)$
  is the azimuthal angle, while $x$, whose range depends on $\eps$,
  resembles the spherical radius.

  For $\eps =-1$, the range of $x$ is $x > 1$ and
\beq                                                          \label{sol-1}
   	\e^{2\gamma} = \biggl(\frac{x-1}{x+1}\biggr)^m,\ \
   	\e^{2\eta} = \biggl(\frac{x^2-1}{x^2-y^2}\biggr)^{m^2};
\eeq
  for $\eps = 0$, $x > 0$, and
\beq                                                          \label{sol0}
	\e^{2\gamma} = \e^{-2m/x},\ \
	\e^{2\eta} = \e^{-m^2 (1-y^2)/x^2};
\eeq
  for $\eps = +1$, $ x\in \R$ and
\beq                                                          \label{sol+}
	\e^{2\gamma} = \e^{-2m \cot^{-1}x},\
	\e^{2\eta} = \biggl(\frac{x^2+y^2}{x^2+1}\biggr)^{m^2}.
\eeq
  All three families have a Schwarzschild asymptotic at large $x$ with
  the mass $M= Lm$. The whole Schwarzschild solution is, however, only
  restored from the family $\eps=-1$, in the case $m=1$ (so that $M =L$),
  and the standard coordinates $r$ and $\theta$ are given by $r = Lx$
  and $y = \cos \theta$.

  In the family $\eps = +1$, the second spatial infinity $x\to -\infty$ is
  also flat, but the Schwarzschild mass is there equal to $-M$. (This feature
  is similar to that of the so-called anti-Fisher \whs, described by
  solutions to the Einstein-scalar equations with a massless phantom scalar
  field \cite{br73, h_ell, ws_book, we-st13}.) The space-time is everywhere
  regular except on the ring $x=0,\ y=0$ whose radius is equal to $\e^{-\pi
  m}L$. The whole configuration is of \wh\ type and has been named
  \cite{BFab97} a ring \wh. The disk $x = 0$ plays the role of a throat, and
  the singularity at its edge ($y=0$) is actually a price paid for \wh\
  existence in vacuum instead of exotic matter. Such ring \whs\ have been
  discusses in a much more general context of D-dimensional Einstein and
  dilaton gravity in \cite{BFab97}.

  It should be stressed that $x > 0$ is not the upper half-space, as could be
  wrongly imagined, but the whole 3-space sheet, connected with another
  sheet, $x < 0$, through the disk $x=0$. Crossing this disk
  (i.e., threading the ring), a hypothetic observer leaves one world and
  enters the other. To return back, he or she should thread the ring once
  again, not necessarily from the same side out of which he/she has appeared
  in this second world.

  Let us now discuss the family $\eps =-1$. In this case, there is no second
  spatial infinity, instead, the limit $x\to 0$ is an attracting singularity
  due to $g_{tt} \to 0$ (except for the case $m=1$ where it is the
  Schwarzschild horizon). However, the radii of circles $x=\const$,
  $y=\const$ infinitely grow as $x\to 0$ if $m >1$. Moreover, the area of
  2D sections $x=\const$ diverges as $x\to 0$ in the case $m >2$.
  Recalling the definition of a throat as a 2D
  surface of minimum area, one can suspect that there must be a throat at
  some finite $x$, contrary to the HV theorem. The situation is illustrated
  by Fig.\,1, showing that the surfaces $x = \const$ do have a minimum area
  if $m > 2$ and do not if $m \leq 2$.

\begin{figure}
\centering
\includegraphics[scale=0.75]{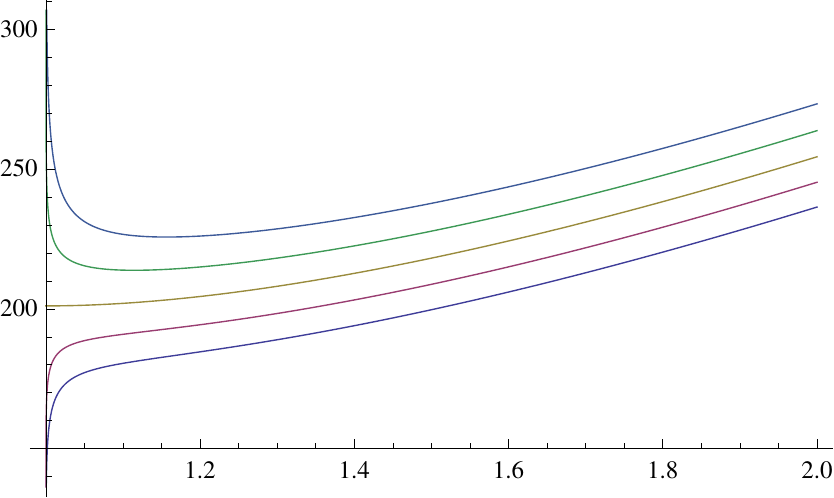}
\caption{\small  Areas $A(x)$ of surfaces $x = \const$ in spaces (\ref{ds-ax})
	with $\eps=-1$ and $m=1.9,\ 1.95,\ 2, 2.05,\ 2.1$ (bottom-up),
	in units $L=1$.}
\end{figure}

  A further study shows, however, that a surface $x=\const$ cannot be a
  throat. Indeed, the minimality condition in terms of the extrinsic
  curvature $K_{ab}$ of a 2-surface in 3D space is that the trace of
  $K_{ab})$ should be zero \cite{HV97}, which, for a surface $x = \const$,
  leads to $(\d/\d x)) \big (\ln |g_{yy}| + \ln |g_{zz}|\big) =0$.
  This condition
  gives a certain function $x(y)$ with a distinct $y$ dependence, contrary
  to the assumption $x = \const$. That is, the assumption that some
  surface $x=\const$ is a throat leads to a contradiction.

\begin{figure}
\centering
\includegraphics[scale=0.8]{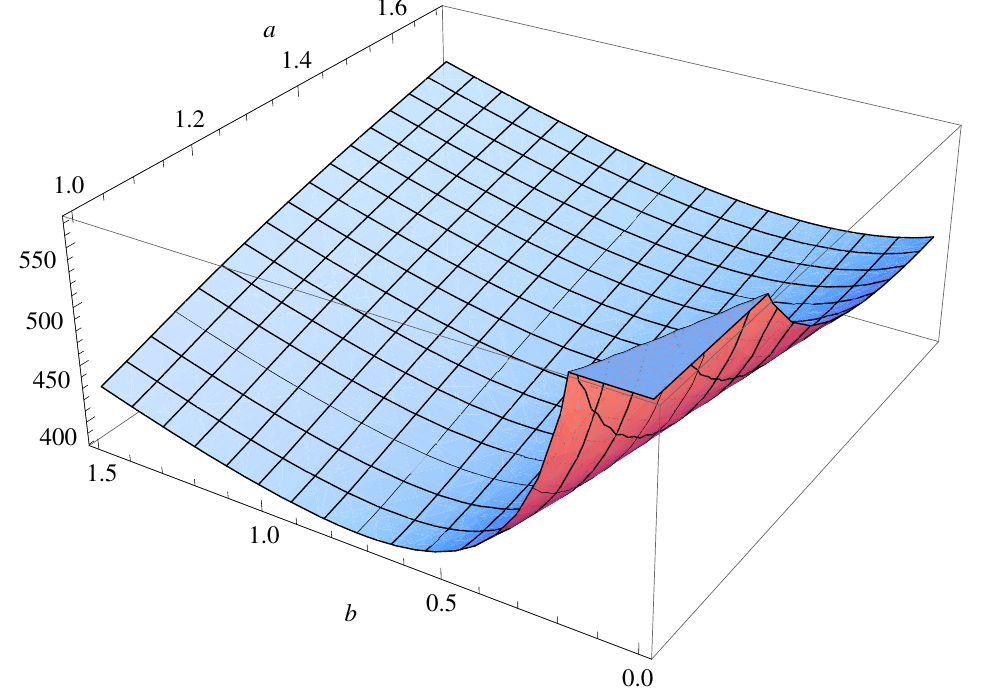}
\caption{\small  Areas $A$ of surfaces $x(y) = a +by^2$ in spaces (\ref{ds-ax})
	$\eps=-1$ and with $m=3$ vs. $a$ and $b$. A minimal surface is absent.
	The section $b=0$, at which $x=\const$ (the right edge), has a minimum
	where $A \approx 478$ in the units used, while at the lowest point
	on the lower edge of the plot (this surface touches the singularity
	$x=1$) $A \approx 382$.}
\end{figure}


  For a more general 2-surface $x = x(y)$, the area is given by a
  functional of $x(y)$,
\beq                                                       \label{A}
     A = 2\pi \int_{-1}^{1} dy \biggl[ g_{\phi\phi}
     	\Big[\Big(\frac{dx}{dy}\Big)^2 g_{xx} + g_{yy}\Big]\biggr]^{1/2},
\eeq
  whose minimum would describe a throat. The corresponding variational
  equation is too complex to try to solve it. Instead, we can use the
  simplest version of Ritz's direct approximation method, namely, to
  consider the family of surfaces
\beq                                                       \label{x(y)}
	x(y) = a + by^2, \ \ \ a = \const \geq 1, \ \ b= \const,
\eeq
  and seek a minimum of $A$ among different pairs of numbers $(a,b)$. A
  numerical calculation leads to the conclusion that no such minimum exists
  for any $m$. The situation is illustrated by Fig.\,2 showing the areas
  $A$ of the surfaces (\ref{x(y)}) for $m = 3$. Consequently, there are no
  throats in the $\eps=-1$ vacuum space-times with the metric (\ref{ds-ax}),
  in full agreement with the HV theorem.

  In conclusion, we would like to mention two more important classes 
  of axially symmetric \whs\ with phantom sources that are not described 
  here, namely,
  rotating axially symmetric \whs\ (see \cite{kuhfittig03, matos09} and 
  references therein) and static multi-\wh\ space-times \cite{sushkov12}.

\end{document}